\documentclass[twocolumn,showpacs,preprintnumbers,amsmath,amssymb]{revtex4}
\usepackage{graphicx}
\usepackage{dcolumn}
\usepackage{bm}
\usepackage{color}

\begin{document}

\title{In situ Ramsey Interferometry and Diffraction Echo with an Ultracold Fermi Gas}

\author{C. Marzok, B. Deh, S. Slama, C. Zimmermann, and Ph.W. Courteille}
\affiliation{Physikalisches Institut, Eberhard-Karls-Universit\"at T\"ubingen,
\\Auf der Morgenstelle 14, D-72076 T\"ubingen, Germany}

\date{\today}

\begin{abstract}
We report on the first observation of Bragg scattering of an
ultracold $^6$Li Fermi gas. We demonstrate a Ramsey-type
matter-wave interferometer based on Bragg diffraction and find
robust signatures of persistent matter wave coherences using an
echo pulse sequence. Because of the Pauli principle, the
diffracted fermions oscillate nearly unperturbed in the trapping
potential for long times beyond 2~s. This suggests extremely long
coherence times. On these timescales, only the presence of a
$^{87}$Rb cloud seems sufficient to induce noticeable
perturbations.
\end{abstract}


\maketitle


Bragg diffraction of cold atoms off moving periodic optical potentials has been
established as a standard tool for coherent atom optics since its
first application to Bose-Einstein condensates (BECs) almost ten years ago \cite{Kozuma99}. Atoms
moving with a wave vector fulfilling the Bragg condition scatter light
from the copropagating into the counterpropagating laser beam and are
accelerated by two photon recoil momenta. Bragg diffraction has
extensively been used in a technique called Bragg spectroscopy
\cite{Stenger99}
serving to measure the dispersion relation and the dynamic
structure factor of BECs and molecular
condensates \cite{Abo-Shaeer05} and to provide signatures of
vortices \cite{Blakie00}. For ultracold fermion
gases, Bragg diffraction is expected to be able to reveal the
pairing mechanism in the BEC-BCS crossover regime by measuring the
dynamic structure factor for both density and spin
\cite{Buechler04}.


Because Bragg diffraction coherently couples two momentum states,
it is frequently used as a beamsplitter for matter waves in
interferometric experiments. Such schemes have been applied to
BECs \cite{Simsarian00}, which bear the advantage of a
macroscopically populated wavefunction yielding a high
interferometric contrast. On the other hand, in BECs the typical
lifetime of coherent superposition states is only on the order of
10\,$\mu$s due to interatomic interactions. For Fermi gases the
matter wave contrast is necessarily limited to single particle
interference, because identical fermions cannot interfere. The
advantage of ultracold spin-polarized Fermi gases is that they are
interaction-free, so that coherent superpositions of motional
states are expected to live very long. Coherence times of hundreds
of milliseconds in superpositions of Wannier-Stark states in a
vertical optical lattice have first shown the superiority of
interaction free Fermi gases for interferometry \cite{Roati04}.


This paper reports the first study of Bragg interferometry with
ultracold fermionic spin polarized $^6$Li atoms inside a magnetic
trap using a moving optical lattice. It turns out that the
momentum distribution after diffraction periodically revives after
multiples of the trap period for more than a hundred times. This
suggests very long coherence times due to the absence of s-wave
interactions. In contrast, the presence of interacting $^{87}$Rb
atoms quickly damps out the revival. With a Ramsey type Bragg
pulse sequence, a matter-wave interferometer can be realized. At
finite temperatures, atoms at different momenta are Doppler
shifted which leads to a detuning of the interferometer for
different momentum classes. A complete Ramsey spectrum can thus be
recorded in a single shot. For long time intervals between the
Ramsey pulses, however, the fringe spacing is beyond the
resolution limit of the imaging system. In contrast to decoherence
as in the case of interacting bosons, which irreversibly destroys
quantum superpositions, dephasing can be reversed by an
appropriate manipulation of the individual atomic phases.
Rephasing of coherent superpositions of atomic excitations has
been demonstrated in spin echo \cite{Hahn50} as well as in photon
echo \cite{Kurnit64}. Phonon echoes have been observed both in
fused silica glasses \cite{Golding76} as well as in cold trapped
atomic clouds \cite{Buchkremer00}. We show that a tailored pulse
sequence consisting of $\frac{\pi}{2}$-$\pi$-$\frac{\pi}{2}$
pulses leads to an echo type revival of the diffraction pattern.
In particular, echo diffraction patterns exhibits very robust
signatures for the degree of coherence of momentum superposition
states, allowing us to detect coherence times of at least
$100~\mu$s in the experiment. Furthermore, simulations show that
such echo type experiments are sensitive to coherence times of
seconds and longer, if the pulses are applied in a stroboscopic
fashion after multiples of the trap oscillation period.




The cooling procedure we use for cooling a mixture of $^6$Li and $^{87}$Rb atoms has been reported in a previous paper \cite{Silber05}. In short, the two species are simultaneously collected by a magneto-optical trap and then
transferred via several intermediate magnetic traps into a
Ioffe-Pritchard type trap, where they are stored in their
respective hyperfine states $|3/2,3/2\rangle$ and $|2,2\rangle$. Two trap configurations are used in the experiment. The \textit{compressed} trap is characterized by the secular frequencies $(\omega_x,\omega_y,\omega_z)/2\pi =(762,762,190)~\text{Hz}$ for
$^6$Li and the magnetic field offset $3.5~$G. For $^{87}$Rb the
trap frequencies are $\sqrt{87/6}$ times lower. The $^{87}$Rb
cloud is selectively cooled by microwave-induced forced
evaporation. The $^6$Li is sympathetically cooled via interspecies thermalization. 
For the experiments described below, we typically reach
temperatures of a few hundred nK with $2..3\cdot10^6$ $^{87}$Rb
atoms and $\sim1.2~\mu$K with $2\cdot10^5$ $^6$Li atoms. The
$^6$Li does not reach the $^{87}$Rb temperature due to the
smallness of the interspecies scattering length \cite{Silber05}.

All experiments, unless stated otherwise, are carried out in a
\textit{decompressed} trap characterized by the secular
frequencies $(\omega_x,\omega_y,\omega_z)/2\pi$
=(236,$\sim$180,141)~Hz for $^6$Li. In this trap, the different
gravitational sagging of the $^{87}$Rb and $^6$Li clouds separates
them in space. Consequently, the $^6$Li cloud does not thermalize
and acquires an anisotropic momentum distribution upon
decompression. For the $z$-axis we measure a momentum width
corresponding to the temperature $T_z\simeq0.9~\mu$K.


Bragg diffraction is performed using two laser beams
counterpropagating along the $z$-axis and tuned $\sim1~$GHz to the
red from the $D_2$ line of $^6$Li. Their frequencies differ by an
amount $\delta=\omega_2-\omega_1=2\hbar q^2/m=2\pi\cdot295~$kHz,
where $q=2\pi/\lambda$, $\lambda=670.977\,$nm and $m$ is the
atomic mass of $^6$Li. The lasers are phase-locked by means of an
electronic feedback circuit. The Bragg beams have intensities
$I_1=I_2=13.7~$..$~132~\text{mW/cm}^2$, corresponding to
one-photon Rabi frequencies of $\Omega_i=\sqrt{6\pi c^2\Gamma
I_i/\hbar\nu^3}$, so that the two-photon Rabi frequency
$\Omega_R=\Omega_1\Omega_2/2\Delta_L$ covers the range $2\pi\,
(47..450)\,$kHz. In most experiments, before applying an in situ
Bragg pulse (or a series of pulses), the $^{87}$Rb atoms are
removed from the trap with a resonant light pulse. After the Bragg
 pulse, we wait for half an oscillation period to rephase the momentum
 distribution before we switch off the trapping field and map the momentum
 distribution after 2 ms of ballistic expansion by absorption imaging.

\begin{figure}[ht]
        \centerline{\scalebox{.5}{\includegraphics{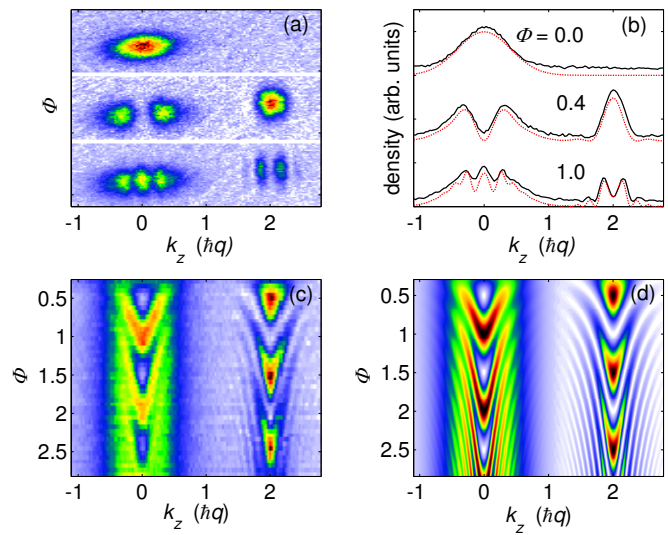}}}
        \caption{(Color online) (a) Time-of-flight absorption images of a $1~\mu$K cold cloud after a single Bragg scattering pulse
            and a $2~$ms ballistic expansion period. From the top to the bottom the Bragg pulse area is varied,
            $\Phi\equiv\Omega_R\tau/2\pi=0$, $0.4$, and $1$.
            The experimental Rabi frequency is $\Omega_R/2\pi=47~$kHz. The tilted shape of the diffracted
            clouds is due to alignment imperfection of the Bragg lasers with respect to the trap's weak axis.
            (b) The solid lines show the integration of the time-of-flight absorption images shown in (a) along the radial direction
            (perpendicular to laser's wavevectors). The dashed lines show simulations of the axial momentum distributions with no free
            parameters.
            (c) False color map of measured momentum distributions for pulse areas varied from 0 to $5\pi$. The radially integrated
            absorption images appear as rows.
            (d) False color map of the calculated momentum distributions using experimental parameters.}
        \label{Fig1}
    \end{figure}

The second image of Fig.~\ref{Fig1}(a) shows a $^6$Li absorption
image taken after application of a single Bragg pulse, whose
length $\tau$ corresponds to $\Omega_R\tau\approx\pi/2$.
 For low Rabi frequencies, only a narrow slice is
cut out of the fermionic momentum distribution.
The position of the slice along the
$z$-axis depends on the detuning of the Bragg lasers from the
two-photon recoil shift, $\Delta=\delta-2\hbar q^2/m$, the
width is due to power broadening by the Rabi frequency
$\Omega_R$.
When the pulse area is increased beyond $\pi/2$, the momentum
distribution acquires a more complicated, axially modulated shape.
This is due to the fact that the various momentum classes of the
atomic cloud are Doppler detuned from the Bragg condition
and therefore experience different effective Rabi flopping frequencies.

The experimental data can be well described by a theoretical
simulation. Fig.~\ref{Fig1}(b,d). In our model we assume that for
every atom the Bragg lasers couple only two discrete momentum
states. The evolution of every momentum class is calculated
individually from the Schr\"odinger equation. The sum of the
results is weighted with the initial momentum distribution. In
cases where the possibility of phase decoherence is considered,
e.g.~by interatomic interactions, we use a two-level Bloch
equation model. As long as the evolution of the momentum
distribution under the action of the trapping potential is not
taken into account, the above treatment only applies to short
pulse sequences, $\tau\ll2\pi/\omega_z$, or to stroboscopic
sequences that are synchronized with the trap period.

\begin{figure}[ht]
       \centerline{\scalebox{.43}{\includegraphics{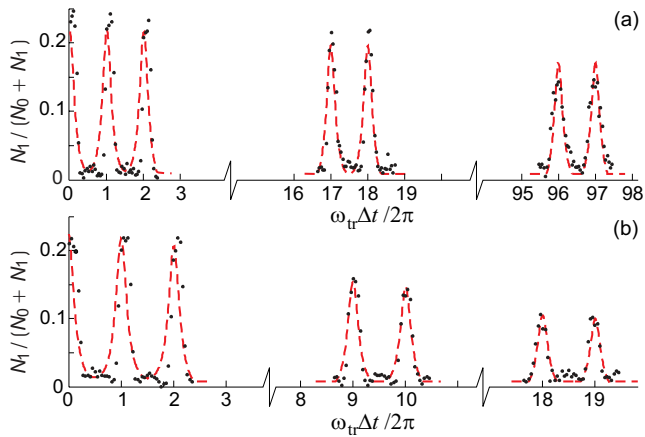}}}
        \caption{(Color online) Rephasing of the momentum distribution in the trap in the absence (a) and in the presence (b) of $^{87}$Rb.
The population of the state $2\hbar q$ normalized to the total number of atoms in states $0\hbar q$ and $2\hbar q$, denoted $N_0$ and $N_1$ respectively. The peaks represent full trap oscillations. The dashed line gives an array of Gaussian functions, whose amplitudes are decreasing exponentially in time. The fits to the data yield the decay times for (a) 2.4~s and for (b) 0.15~s. The oscillation frequency was $\omega_z/2\pi=142.51~$Hz and the Rabi frequency $\Omega_R/2\pi=47~$kHz.}
        \label{Fig2}
    \end{figure}

To study the impact of both the trap and of collisions on the
momentum distribution, we apply a single $\pi$-pulse, let the
atoms oscillate in the trap for a long time $t\gg2\pi/\Omega_R$
and then image the cloud's momentum distribution. This experiment
has been performed in the compressed trap, because in this
configuration there is spatial overlap between $^6$Li and
$^{87}$Rb clouds. By counting the numbers of atoms in a restricted
area of momentum space corresponding to the first-order
Bragg-diffracted atoms, we obtain the curves shown in
Fig.~\ref{Fig2}(a) for the case of no $^{87}$Rb inside the trap.
Apparently, the atoms oscillate more than a hundred times with
only small damping corresponding to an exponential decay time of
2.4~s. This extremely long dephasing time impressively
demonstrates that diffusion in momentum space is very slow
\cite{Note01}. As collisions, which are forbidden by the Pauli
principle, are absent, momentum dephasing is mainly due to
residual anharmonicities of the trapping potential. In this
respect the Fermi gas efficiently emulates the dynamics of an
ideal gas of non-interacting particles.

We tested the impact of collisions by keeping the $^{87}$Rb in the
trap. For this case, Fig.~\ref{Fig2}(b) shows a dramatic reduction
of the decay time for the $^6$Li cloud to 0.15~s. This value
agrees well with the inverse collision rate estimated for the
densities and temperatures specified above,
$\gamma_{coll}^{-1}\simeq0.16~$s \cite{Silber05,Marzok07}.
Thus, for the mixture, we expect coherent states to decay on this time
scale.
The existence and the lifetime of such states can be tested by interferometric
experiments. Furthermore, they justify stroboscopic timing for
absorption imaging and for interferometry.

\begin{figure}[ht]
        \centerline{\scalebox{.5}{\includegraphics{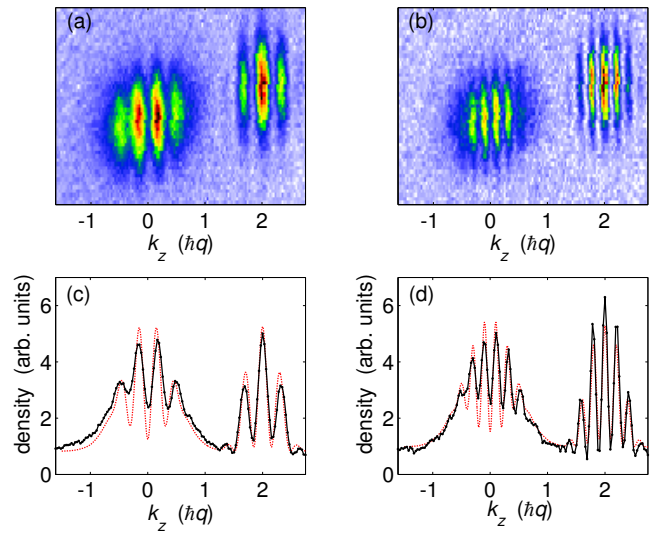}}}
        \caption{(Color online) (a,b) Time-of-flight absorption images taken after a Ramsey pulse sequence and (c,d) their radial integrations (black solid line).
            The times between the Ramsey pulse are (a,c) $6~\mu$s and (b,d) $12~\mu$s. The theoretical curves (red dotted line) in (b,d) are fits based on
            our model. $\omega_z/2\pi=142.51~$Hz and $\Omega_R/2\pi=57~$kHz are as in Fig.~\ref{Fig1}.}
        \label{Fig3}
    \end{figure}

To check the lifetime of coherent superpositions of momentum
states, we apply a Ramsey sequence of two $\pi/2$ Bragg pulses
separated by a variable short time interval $\Delta
t\ll2\pi/\omega_z$. The inhomogeneous momentum distribution of the
atoms corresponds to a Doppler-shift, which detunes the atoms from
the resonant Bragg condition. Therefore, the atoms accumulate
different phases during their free evolution in the trap, which
allows to observe a full Ramsey spectrum in a single shot, as
shown in Fig.~\ref{Fig3}. While we do observe Ramsey fringes for
time intervals up to $\Delta t\simeq32~\mu$s, for longer times,
the fringe spacing is below the resolution limit of the imaging
system. Lower temperatures improve the fringe contrast without
influencing the fringe spacing. The resolution can also be
somewhat enhanced by longer times of flight at the expense of
contrast.


More robust signatures are needed to find out, whether
coherence is preserved for times exceeding $32~\mu$s. A
well-known method to rephase the momentum distribution and to
obtain Doppler-free signals is echo interferometry
\cite{Andersen03}. This method divides the free
evolution time in a Ramsey experiment in two periods separated by a
$\pi$-pulse. The $\pi$-pulse inverts the accumulated phase delay of the different atomic momentum classes so that the states rephase during the time interval after the $\pi$-pulse. Thus, all momentum classes contribute to the signal.
    \begin{figure}[ht]
        \centerline{\scalebox{.5}{\includegraphics{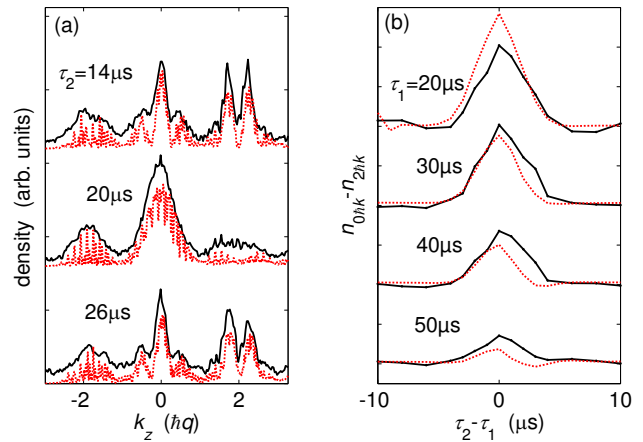}}}
        \caption{(Color online) (a) Radially summed absorption images (black solid line) taken after an echo sequence  $\frac{\pi}{2}-\tau_1-\pi-\tau_2-\frac{\pi}{2}$
        for $\tau_1=20\,\mu$s. Also shown are simulated momentum distributions calculated with experimental parameters (red dotted line).
        (b) Difference of the measured (black solid line) and simulated (red dotted line) relative populations of
        the states $k_z=0\hbar q$ and $k_z=2\hbar q$ for varying $\tau_2$ with different fixed values of
        $\tau_1$. The echo signal washes out for $\tau_1$-times larger than 50\,$\mu$s.}
        \label{Fig4}
    \end{figure}

The momentum distributions observed after an echo sequence in the
regime of \textit{small} Rabi frequencies $\Omega_R\ll\hbar k_{\rm
th}^2/2 m$ exhibit a central dip for atoms near $2\hbar q$, whose
width depends on $\Omega_R$ (data not shown). Theoretical
simulations show that low-energy atomic momentum classes, which
are within the area of this dip are efficiently rephased by the
echo pulse scheme and the $2\hbar q$ state is depopulated by the
second $\frac{\pi}{2}$-pulse. For higher
momenta, the Doppler-shift causes the pulse areas to deviate considerably from $\frac{\pi}{2}$ which corrupts the echo rephasing scheme.
Nevertheless, simulations reveal that the width of the dip is very
stable even after long times $\Delta t\gg2\pi/\omega_z$. In
contrast, in the presence of decoherence, the dip vanishes. The
appearance of the dip in the momentum profile thus represents a
robust signature for coherence, preferable to Ramsey fringes as it
overcomes limits in momentum resolution.

For \textit{higher} Rabi frequencies, more atoms contribute to the
echo signal, thus increasing the signal-to-noise ratio. For
$\Omega_R\geq\hbar k_{\rm th}^2/2 m$ the whole population of the
momentum states has to be monitored. In Fig.~\ref{Fig4},
diffraction echo is demonstrated for $\Omega_R/2\pi=220\,$kHz. A
$\frac{\pi}{2}$-$\pi$-$\frac{\pi}{2}$ pulse sequence with delay
times of $\tau_1$ and $\tau_2$ between the pulses has been
applied. In Fig.~\ref{Fig4}(a) diffraction echo manifests itself
in an almost complete revival of the $k_z=0\hbar q$ momentum state
when $\tau_1=\tau_2=20~\mu$s. As is apparent in the data, the
$k_z=-2\hbar q$ state is also slighty populated as one approaches
the crossover regime from Bragg scattering to Kapitza-Dirac
scattering. An extended model including higher momentum states as
well as trap oscillations successfully reproduces the measured
momentum distributions (red dotted line). The initial spatial
distribution is also taken into account for each $k_z$
individually. The spikes in the model curve arise from the limited
number of initial space coordinates used for averaging due to
limitations in available computing power. Subtracting the
population of state $2\hbar q$ from state $0\hbar q$ shows the
quantitative effect of diffraction echo [Fig. \ref{Fig4}(b)]. The
echo signal fades out after $100~\mu$s. This is an effect of the
spatial distribution and does not need decoherence for an
explanation. The extended model predicts a revival of the echo
signal for $\tau_1=\tau_2=T_{\rm trap}=7.092\,$ms, where $T_{\rm
trap}$ is the trap period. Measurements have not shown such a
revival, although we expect coherence to persist on this
timescale. Possible technical reasons which could account for an
incomplete refocussing of the atomic phases could be fluctuating
Ramsey pulse lengths, power or frequency fluctuations of the Bragg
lasers, or alignment imperfections of the Bragg beams with respect
to the $z$-axis.


In conclusion, we studied the coherence lifetime of an ultracold
Fermi gas via \textit{in situ} Bragg interferometry.
We demonstrated that the method of diffraction echo is capable of producing robust signatures of coherence, but is today limited by practical imperfections of the
experimental setup. Simulations reveal, that a stroboscopic diffraction echo scheme synchronized with the trap period is suitable for observing the expected long coherence times.

We also used Bragg spectroscopy for mapping the momentum
distribution of the $^6$Li cloud. Details will be reported
elsewhere. Differently from time-of-flight absorption imaging the
resolution of this method is not limited by the imaging system,
which may allow to resolve small scale features in the momentum
distribution with high precision. Since the temperatures in our experiment are
not deep in the Fermi regime, the integrated momentum distribution does not
deviate much from that of an ultracold classical gas
\cite{DeMarco98}, such that we do not observe clear signatures of Fermi
pressure.

Now shown to be applicable to fermions, Bragg spectroscopy may
prove useful for probing distinct features of the dispersion
relation such as rotons \cite{Steinhauer04} or long-range
many-body correlations such as Cooper-pairing or fermionic
condensation \cite{Bruun06}. In the future we aim to study the
dynamical properties of heteronuclear spin mixtures with tunable
interactions by performing Bragg scattering in the vicinity of one
of the recently found heteronuclear Feshbach resonances
\cite{Deh08}. Further, studies in the crossover regime from Bragg
scattering to superradiance with fermions are planned
\cite{Ketterle01}.

\bigskip
This work has been supported by the Deutsche Forschungsgemeinschaft (DFG) within the Schwerpunktprogramm SPP1116. We acknowlegde helpful discussions with Andreas G\"unther and Martin Zwierlein.

\end{document}